\documentclass[aps,prb,onecolumn,showpacs]{revtex4-1}
\usepackage{amssymb}

\usepackage{graphicx}
\usepackage{xcolor}

\begin{document}

\title{Spin polarization and orbital effects in superconductor-ferromagnet
structures}
\author{A. F. Volkov}
\email{volkov@tp3.rub.de}
\affiliation{Theoretische Physik III,
Ruhr-Universit\"{a}t Bochum, D-44780 Bochum, Germany
 }
\author{F. S. Bergeret}
\email{fs.bergeret@csic.es}
\affiliation{Centro de F\'isica de Materiales (CFM-MPC), Centro Mixto CSIC-UPV/EHU,
	Manuel de Lardizabal 5, E-20018 San Sebastian, Spain}
\affiliation{Donostia International Physics Center (DIPC), Manuel de Lardizabal 4,
	E-20018 San Sebastian, Spain}

\author{K. B. Efetov}
\affiliation{Theoretische Physik III,\\
	Ruhr-Universit\"{a}t Bochum, D-44780 Bochum, Germany.}
\date{\today }

\begin{abstract}
We study theoretically spontaneous currents and magnetic field induced in a
superconductor-ferromagnet (S-F) bilayer due to direct and inverse proximity
effects. The induced currents {are Meissner currents that appear even in the
absence of an external magnetic field due to the magnetic moment in the
ferromagnet }and {to the  magnetization } in the superconductor . The
latter is induced by the inverse proximity effect over a distance of the
order of the superconducting correlation length $\xi _{S}$. On the other
hand the magnetic induction $B$, caused by Meissner currents, penetrates the
S film over the London length $\lambda _{S}$. Even though $\lambda _{S}$
usually exceeds considerably the correlation length, the amplitude and sign
of $B$ at distances much larger than $\xi _{S}$ depends crucially on the
strength of the exchange energy in the ferromagnet and on the magnetic
moment induced in the in the S layer.
\end{abstract}

\maketitle
\date{\today}

\section{Introduction}

Besides the orbital effects, it is well known that conventional
superconducting pairing is also {suppressed} by a magnetic field when acting
on the spins of electrons via the Zeeman interaction. Whereas superconducting
correlations couple pairs of electrons in a singlet state (Cooper pairs),
the Zeeman interaction tends to align both spins in the direction parallel
to the magnetic field. These two antagonistic tendencies can nevertheless
coexist when the Zeeman energy is small enough in comparison to the
superconducting gap. This coexistence implies the appearance of Cooper pairs
in a triplet state. This situation occurs, for example, in thin
superconducting films (S films) in the presence of an in-plane magnetic
field, and also in superconductor-ferromagnet (S-F) heterostructure in which
Cooper pairs from the S layer can penetrate into the ferromagnet where the
intrinsic exchange field $J$ of the F acts on the spins of electrons (see
review articles \cite%
{GolubovRMP04,BuzdinRev05,BVErev05,EschrigRev11,LinderRev15,LinderBalRev17}).

Leakage of Cooper pairs from S to F is the so called proximity effect. The
wave function of the Cooper pairs penetrating into the F region {\ with a
uniform magnetic moment} $\mathbf{M}$ contains not only the singlet but also
the triplet component with zero spin  projection  onto  the vector $%
\mathbf{M}$. At the same time, provided the S-F interface is transparent
enough, these triplet pairs can leak into the superconductor,  inverse
proximity effect,  and a finite magnetic moment in S appears\cite%
{BVE04,BVEepl04,footnoteM}.

The {size} of this spin-polarized region within the superconductor is of the
order of the superconducting correlation length, which in the diffusive
limit is given by $\xi _{S}\approx \sqrt{D_{S}/2\pi T_{c}}$. {The magnetic
moment }$\mathbf{M}_{S}$ induced in S, has a direction opposite to the
{magnetization vector} $\mathbf{M}_{F}$ in the F layer. Under certain
conditions the total magnetic moment in the S region compensates the total
magnetic moment of the F film resulting in a full spin screening\cite%
{BVE04,BVEepl04,foot} In the ballistic case, the induced magnetization $%
\mathbf{M}_{S}(x)$ may spatially change sign \cite{Bergeret05,Kharitonov06}.

These predictions for the inverse proximity effect have eventually been
confirmed experimentally \cite{Garifullin09,Kapitulnik09}. However,
quantitative interpretation of the experimental {results }is quite subtle%
\cite{diBernardo,Flokstra}, since the magnetic field arising in S is caused
not only by the induced magnetization $\mathbf{M}_{S}$ but also by Meissner
currents that arise in the S/F structure\cite{PRB01,BVEepl04,Buzdin18}. For
this reason a detailed understanding of the inverse proximity effect is a
key issue for interpretation of experimental data on S/F structures.

In this work we study the proximity effect in S-F structures {taking into
account explicitly} the generated spontaneous currents. This topic was first
addressed by the authors in 2004 \cite{BVEepl04} and more recently in Ref. %
\cite{Buzdin18}. These two works predict a magnetic induction $B_{s}(x)$
induced in the S layer which penetrates over the London penetration depth $%
\lambda _{S}$. {The authors of Ref.\cite{BVEepl04} focus on the magnetic
field caused by }$\mathbf{M}_{S}(x)$ {and estimated the orbital effects.
They showed that the spin polarization effects are stronger than those
related to the Meissner currents screening the magnetic moment }$\mathbf{M}%
_{F}${. In Ref.\cite{Buzdin18} the orbital effects were studied in more
detail, but the inverse proximity effect was completely neglected}. Since in
dirty superconducting films $\lambda _{S}$ is usually larger than the
coherence length $\xi _{S}$ characterizing penetration of a magnetic moment $%
M_{S}$ into S,  it might look at first glance as if the magnetic field measured in a
superconducting film with a large  thickness  $d_{S}$ ($\xi _{S}\ll d_{S}\lesssim
\lambda _{S}$), could not be affected by the magnetic moment $M_{S}$
localized close to the S-F interface. In contrast to this scenario, we
demonstrate here that for a correct interpretation of the experimental data
one does need to take into account the magnetic moment in the S layer
induced by the inverse proximity effect.

To be specific, we show that the spatial dependence of the magnetic
induction $B_{S}(x)$ induced in the S region consists of a long-range
component $B_{l-r}(x)$ which decreases over the  London penetration depth $\lambda
_{S}$,  and of a short-range component $B_{s-r}$ caused by the induced
magnetization which  decays over the superconducting coherence length $\sim
\xi _{S}$. The magnetic inductance, $B_{S}$,  in a thick superconducting film\ (%
$\xi _{S}\ll d_{S}\lesssim \lambda _{S}$) has thus the form:
\begin{equation}
B_{S}(x)\text{ }=B_{l-r}(0)\exp (x/\lambda _{S})+B_{s-r}(x)\;,  \label{I4}
\end{equation}%
At large distances from the S/F interface, $|x|\gg \xi _{S}$ , $B_{S}(x)$ is mainly  
determined by the long-range term $B_{l-r}(x)$. Its
amplitude consists of two contributions:
\begin{equation}
\mathbf{B}_{l-r}\mathbf{(0)=B}_{orb}\mathbf{(0)+B}_{sp}\mathbf{(0)\;.}
\label{I4a}
\end{equation}%
{The first term is the contribution from the spontaneous Meissner currents
(orbital effects) and equals }%
\begin{equation}
B_{orb}(0)=-4\pi M_{0}\theta _{F}^{2}/2\text{ }\;,  \label{I5}
\end{equation}%
where $\theta _{F}=(d_{F}/\lambda _{F})$, $d_{F}$ and $\lambda _{F}$ are the
thickness of the F layer and the London penetration depth in the
ferromagnet, respectively. This expression coincides with the result for the
magnetic induction obtained in Ref. \cite{Buzdin18}.  One of our main findings below, is  that
there is an additional contribution to the magnetic induction, the term $B_{sp}$ in
Eq. (\ref{I4a}). This contribution is  caused by the inverse proximity effect it was  neglected in
Ref. \cite{Buzdin18}. For a wide range of parameters this contribution due
to spin polarization near the S/F interface is much larger than that due to
orbital effects, \textit{i.e.} $B_{sp}(0)\gg $ $B_{orb}(0)$. Moreover, this
contribution might be crucial in determining the sign of the magnetic
induction in the S layer since, as we show below,  $B_{sp}(0)$ and $B_{orb}(0)$ have different signs.
Moreover, the relative magnitude between these two contributions depends on the exchange field $J$ in the F layer. 
The contribution $B_{sp}(0)$ due to spin polarization in S
can be neglected only in case of F film with sufficiently large exchange
energy $J$.



In the next sections we investigate the spatial distribution of the Meissner
currents $j_{S}(x)$ and the fields $B(x)$, $H(x)=B(x)-4\pi M(x)$. Our main
findings are the following: (i) Eq.(\ref{I5}) describes the orbital effect only in the
case of rather large exchange energy $J$. { However, in this
case the induced field $B_{orb}(0)$ is small since the inverse London
penetration depth $\lambda _{F}^{-1}\propto J^{-4}$ is small;} (ii) In the
full screening case, both short- and long-range components in Eq.(\ref{I4a})
are determined by spin polarization effects; (iii) Meissner currents in the
S region change sign at some point $x_{0}\sim\xi
_{S}$ away from the S/F interface; (iv) the total Meissner currents in the F (or S) film calculated with
or without account for the spin screening effect may have opposite
directions, and (v) in the case of an out-of-plane magnetization of the F
layer no spontaneous currents, and hence no magnetic induction, are induced.

The article is organized as follows. In the next section we consider a
diffusive S/F bilayer and derive the expressions for the magnetic moment $%
M_{S}$ induced by the spin polarization. Although this have been presented
in our earlier publications \cite{BVE04,BVEepl04}, for completeness and to
set the notation, we re-derive it here. In section III we solve the magnetostatic equations for the vector potential and 
 find the  spatial
distribution of the spontaneous supercurrents $j$\ in the system which arise
in the absence of an external magnetic field. In particular we show that the
current density $j_{S}$\ in the S film can change its  sign near the S/F interface.
In the last section we summarize our results.

\section{Proximity effect in S-F Heterostructure}

In this section we study the proximity effect in {an} S-F strcuture.
We assume the diffusive limit, such that the conditions $\Delta \tau \ll 1$
and $J\tau \ll 1$ are satisfied, where $\tau $ is the momentum relaxation
time. The presence of a spin-dependent term in the F region means that the
condensate induced in this layer consists of a singlet and triplet
component. In turn, triplet Cooper pairs may penetrate into the S region and
induce a finite spin polarization. To describe this processes in a diffusive
system, we use the quasiclassical Green's functions (GF) $\hat{g}_{S}(\omega
)$ \cite{Kopnin,Footnote1} and the generalized Usadel equation \cite%
{BVErev05,BSVH2018}.

\begin{figure}[t]
\includegraphics[scale=0.3]{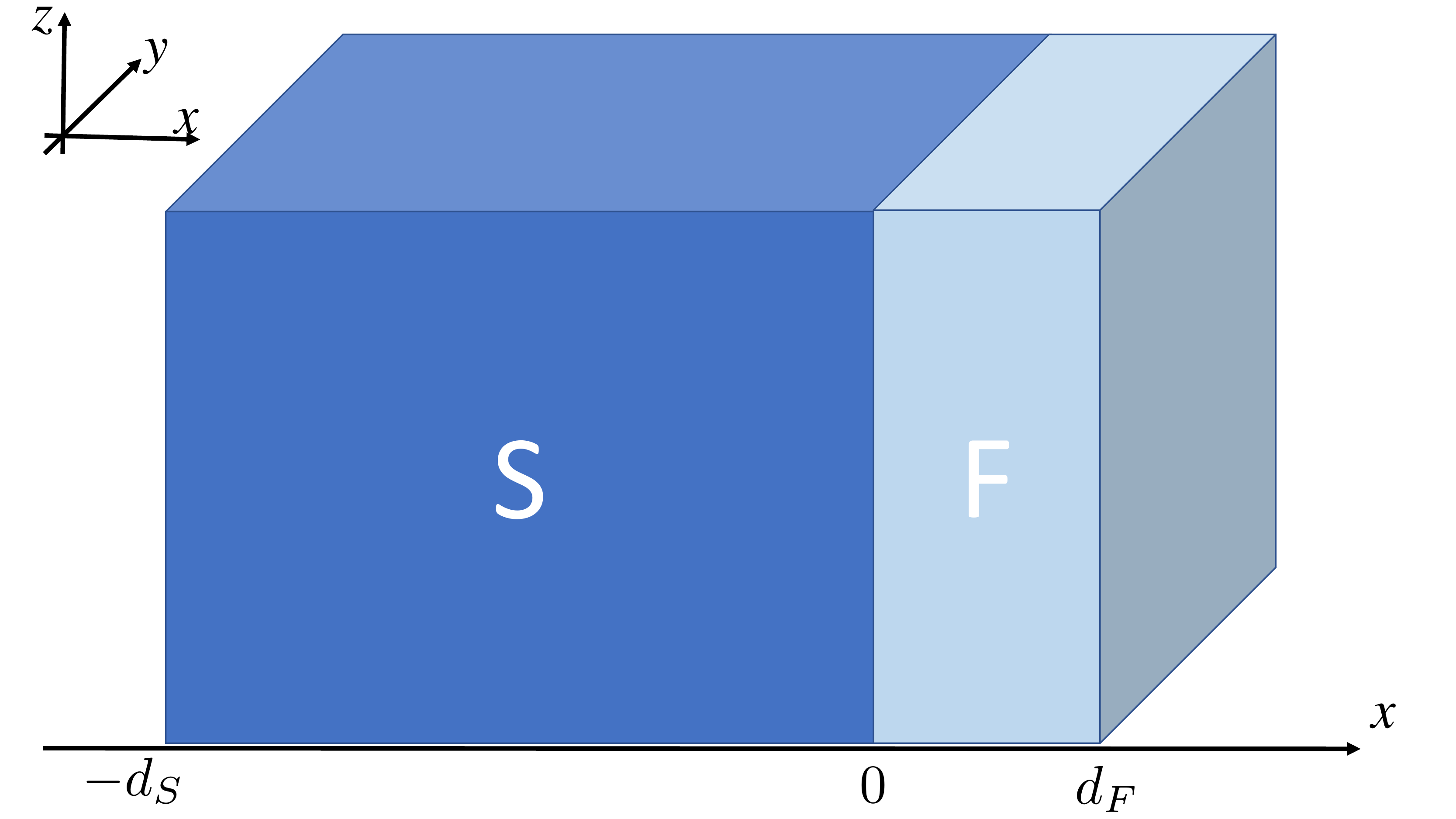}
\caption{The S-F structure under consideration. We assume that the
ferromagnet has an homogeneous magnetization in $z$-direction. }
\label{fig:geom}
\end{figure}

Specifically, we consider the structure shown in Fig. \ref{fig:geom}. It
consists of a ferromagnetic layer of thickness $d_{F}$ and a superconducting
layer of thickness $d_{S}$. In the absence of proximity effect\ the Green
function in S corresponds to the the bulk BCS matrix Green function $\hat{g}%
_{S}(\omega )$ which has the form
\begin{equation}
\hat{g}_{S}(\omega )=G_{S}\hat{\tau}_{3}+F_{S}\hat{\tau}_{1}  \label{1}
\end{equation}%
where $\hat{\tau}_{1,3}$ are the Pauli matrices operating in the
particle-hole space and $G_{S}=\omega /\zeta _{\omega }=(\omega /\Delta
)F_{S}$, $\zeta _{\omega }=\sqrt{\omega ^{2}+\Delta ^{2}}.$ Here $\omega $
is the fermionic Matsubara frequency.

In the presence of an exchange field $J$, the quasiclassical Green's
function $\hat{g}$ maintains its structure in the particle-hole space but
its components are matrices in the spin-space. We consider here only a
mono-domain ferromagnet with an homogenous $J$ and therefore the general
form of $\hat{g}$ in S and F is
\begin{equation}
\hat{g}_{a}(\omega )=(g_{a0}\hat{1}+g_{a3}\hat{\sigma}_{3})\hat{\tau}%
_{3}+(f_{a0}\hat{1}+f_{a3}\hat{\sigma}_{3})\hat{\tau}_{1}  \label{2}
\end{equation}%
where $\hat{\sigma}_{3}$ is the third Pauli matrix in the spin space, and
the index $a$ means $a=S,F$. %
In Eq.(\ref{2}) the terms proportional to $\tau _{3}$ are the normal Green
funcions (GF)which determine the electronic charge and spin densities.
The terms proportional to $\tau _{1}$ are the anomalous GF
describing the singlet and zero-spin projection triplet components of the
condensate. Without losing generality we assume that the exchange field $J$
points in $z$-direction, $J=J\hat{\mathbf{z}}$.

The GFs can be calculated by solving the Usadel equation complemented with
proper boundary conditions (see Appendix \ref{app_usadel} for detail). The
GF calculated in this way determine the current and electron magnetization
density, $M=M\hat{\mathbf{z}}$ as follows
\begin{equation}
\mathbf{j}_{a}{=}\frac{1}{4}\sigma (2\pi i)T\text{Tr}\sum_{\omega }(\tau _{3}%
\hat{g}_{a}\mathbf{\nabla }\hat{g}_{a})\text{, }  \label{6a}
\end{equation}

\begin{equation}
M_{a}(x)=M_{0}(x)+\frac{1}{4}(2\pi i)T\mu _{B}\nu \text{Tr}\sum_{\omega
}(\tau _{3}\sigma _{3}\hat{g}_{a})\; ,  \label{6b}
\end{equation}%
here $\mu _{B}$ and $\nu $ are an effective Bohr magneton and the normal
density of states at the Fermi level respectively. $M_{0}$ is the
magnetization in the normal state which is finite, and spatially
homogeneous, only in the F layer.

Clearly the matrix $\hat{g}$ defined in Eq.(\ref{2}) is diagonal in the spin
space. This simplifies the calculation of $\hat{g}$ since the equations for
up and down spins decouple from each other.
In other words, one can write the GF as $\hat{g}_{a\pm }=g_{a\pm }\tau
_{3}+f_{a\pm }\tau _{1}$ and solve the problem independently for $\pm $
spins. Qualitatively, due to conventional proximity effect, a spin dependent
condensate function $\hat{f}_{F\pm }$ is induced in the F layer. Such
spin-polarized condensate can penetrate back into the S region inducing a
local magnetic moment described by the corrections to the GF $\delta \hat{g}%
_{S\pm }$ defined as $\delta\hat{g}_{S\pm }=\hat{g}_{S\pm }-\hat{g}_{S}$.
All these functions can be obtained from the Usadel equation, as explained 
 in Appendix \ref{app_usadel}. 
 
In order to solve the problem analytically we assume that the F film is thin enough 
\textbf{,} $d_{F}\ll \sqrt{D_{F}/J}$, and therefore the matrix $\hat{g}_{F}$
can be considered almost constant in space. We also assume that the S-F interface
has a finite interface resistance per unit area, $R_{b}$. This allows us to
use the Kupriyanov-Lukichev boundary condition (\ref{A5}). Then we can
integrate spaatially the Usadel equation, Eq. (\ref{A4}), in the F region to obtain 
following algebraic equation for $\hat{g}_{F\pm }$:
\begin{equation}
\lbrack \tilde{\omega}_{\pm }\hat{\tau}_{3}+\tilde\Delta \hat{\tau}_{1},%
\hat{g}_{F\pm }]=0  \label{7a}
\end{equation}%
where $\tilde{\omega}_{\pm }=\omega +\epsilon _{bF}G_{S}\pm iJ$, $\tilde{%
\Delta}=\epsilon _{bF}F_{S}$, $\epsilon _{bF}=D_{F}/(R_{b}\sigma _{F}d_{F})$%
, and $\sigma _{F}$ is the conductivity of the F layer. Equation (\ref{7a})
has to be solved together with the normalisation condition $\hat{g}_{\pm
}^{2}=1$. The solution has the same structure as the bulk BCS solution with
renormalized $\omega $ and $\Delta $. \textit{cf.} Eq, (\ref{1})
\begin{equation}
\hat{g}_{F\pm }=(\tilde{\omega}_{\pm }\hat{\tau}_{3}+\tilde{\Delta}\hat{\tau}%
_{1})/\tilde{\zeta}_{\omega \pm },  \label{8}
\end{equation}%
where $\tilde{\zeta}_{\omega \pm }=\sqrt{\tilde{\omega}_{\pm }^{2}+\tilde{%
\Delta}^{2}}$.

\begin{figure}[t]
\includegraphics[scale=0.3]{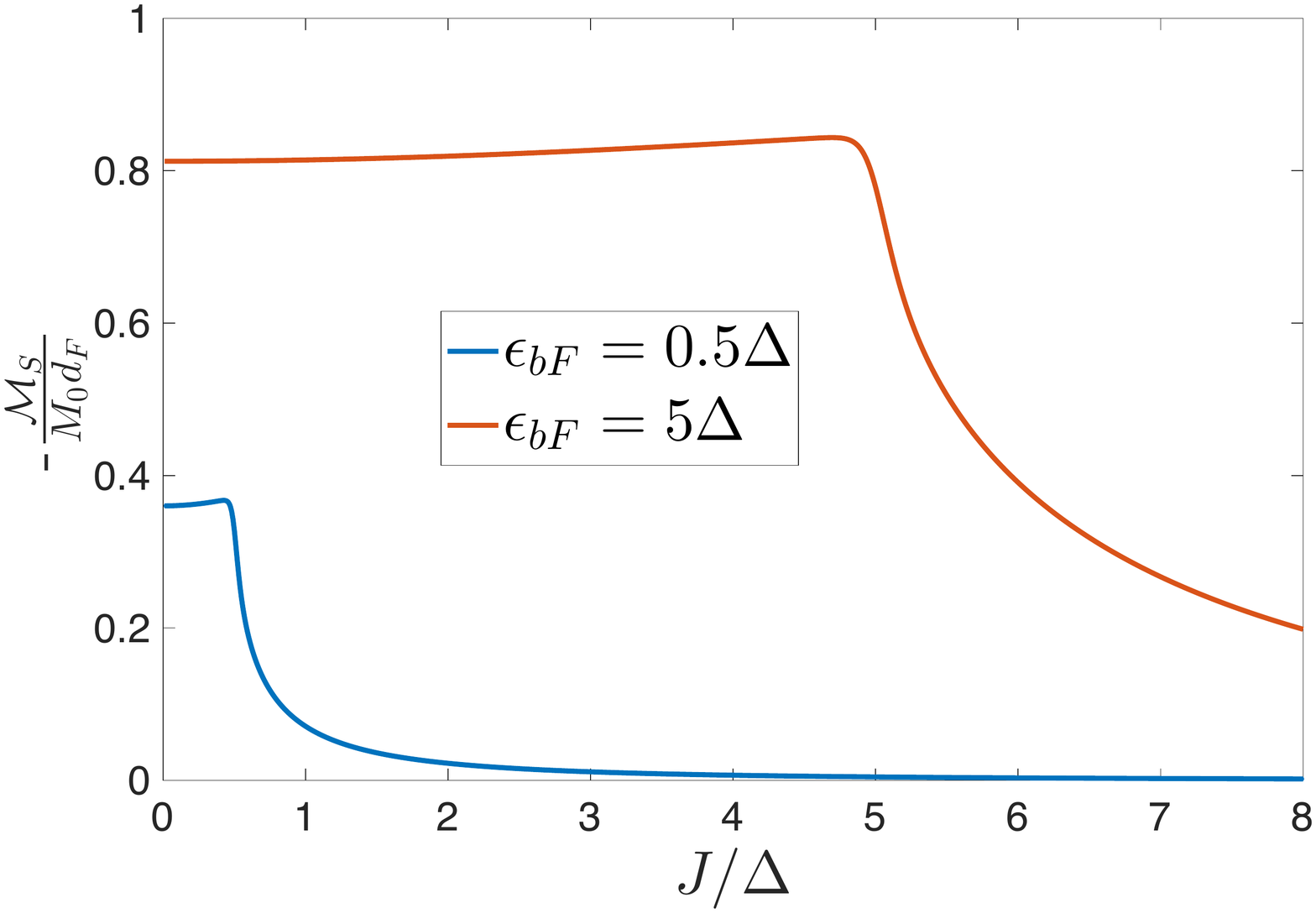}
\caption{The total magnetization induced in the S layer as a function of $J$
for two different values of $\protect\epsilon_{bF}$, and $d_{F}=\protect\xi %
_{S}$, $d_{S}=2\protect\xi _{S}$, and $\protect\kappa _{Fb}=5\protect\kappa %
_{Sb}=0.5\protect\xi _{S}^{-1}$.}
\label{figtotmagn}
\end{figure}

On the superconducting  side of the interface, the GF  are modified due
to the inverse proximity effect. Provided the transmission of the S/F
interface is finite, a correction $\delta \hat{g}_{S}$ to the BCS Green's
functions arises in the S film. We assume that the elements of the matrix $%
\delta \hat{g}_{S}$ are smal\.{l}: $|\delta \hat{g}_{S}|\ll 1$. Then, in the
leading order approximation we obtain $g_{S\pm }\approx G_{S}+\delta g_{s\pm
}$ and $f_{S\pm }\approx F_{S}+\delta f_{S\pm }.$ So, the magnetisation
density induced in the S film is given by
\begin{eqnarray}
M_{S}(x) &=&2i\pi T\mu _{B}\nu _{S}\sum_{\omega \geq 0}g_{S3}(\omega ,0)\exp
(x\kappa _{\omega })\equiv  \label{13a} \\
&\equiv &-\sum_{\omega \geq 0}m_{S}(\omega )\exp (x\kappa _{\omega })\text{.}
\nonumber
\end{eqnarray}%
where $m_{S}(\omega )\equiv -2\pi iT\mu _{B}\nu _{S}g_{S3}^{(S)}(\omega ,0)$%
, and $\kappa _{\omega }^{2}=2\sqrt{\omega ^{2}+\Delta ^{2}}/D_{S}.$ The
function $g_{S3}(\omega ,0)$ is defined in Eq. (\ref{2}) and explicitly
given in the appendix Eq.(\ref{A13}). The total magnetic moment induced in
the superconductor\textit{\ }$\mathcal{M}_{S}$ is obtained by integrating
the previous expression in the interval $-\infty <x<0$
\begin{equation}
\mathcal{M}_{S}=\int_{-\infty }^{0}dxM_{z}(x)=-\sum_{\omega \geq 0}\frac{%
m_{S}(\omega )}{\kappa _{\omega }}.  \label{14}
\end{equation}%
Using Eqs.(\ref{2},\ref{A13}), we reduce Eq.(\ref{14}) to the form

\begin{equation}
\mathcal{M}_{S}=-M_{0}d_{F}(\epsilon _{bF}/J)(2\pi T)\text{Im}\sum_{\omega
\geq 0}\frac{\Delta ^{2}}{\zeta _{\omega }^{3}}\frac{\omega +iJ}{\tilde{\zeta%
}_{\omega }}  \label{14b}
\end{equation}%
{where the functions $\zeta _{\omega }$\ and $\tilde{\zeta}_{\omega }$%
\ are defined in Eqs.(\ref{1},\ref{8}). In Fig. \ref{figtotmagn} we
represent the dependence of the normalized total magnetic moment in the S
region $M_{S}$\ on the exchange energy $J$ for two values of $
\epsilon _{bF}$: $\epsilon _{bF}> \Delta$ and $\epsilon _{bF}<\Delta$. One
can clearly see a kink at $J\cong \epsilon _{bF}$. When $\epsilon
_{bF}<\Delta$, the characteristic energy $\epsilon _{bF}$\ describes the
subgap induced in the F film due to proximit effect. In this case {the
induced magnetization is small and there is no full screening}. In contrast,
in the limit $\epsilon _{bF}>\Delta$ an almost full screening takes place
provided $J<\epsilon_{bF}$}

One can analytically calculate the total magnetisation $M_{S}$\ in the
superconducting region in the limit $\Delta (T),T\ll J\ll D_{F}/(R_{b}\sigma
_{F}d_{F})\equiv \epsilon _{bF}$. {This condition combined with Eq.(%
\ref{A13c}) can be written as}
\begin{eqnarray}
\rho _{S}\xi _{S} &\ll &R_{b}\ll \rho _{F}d_{F}\frac{E_{Th}}{\Delta }
\label{15} \\
J &\ll &E_{Th}\equiv D_{F}/d_{F}^{2}
\end{eqnarray}
In this limit one obtains $\tilde{\omega}_{\pm }\approx \epsilon
_{bF}(G_{S}+i\tilde{J})$\ and $\tilde{\zeta}_{\omega +}\approx \epsilon
_{bF}(1+i\tilde{J}G_{S})$, where $\tilde{J}=J/\epsilon _{bF}$. The term ($%
m_{S}(\omega )/\kappa _{\omega }$) in Eq.(\ref{14}) is approximately equal
to
\begin{equation}
\frac{m_{S}(\omega )}{\kappa _{\omega }}\cong (2\pi T\mu _{B}\nu _{S})\frac{%
2\kappa _{Sb}}{\kappa _{\omega }^{2}}F_{S}^{2}\tilde{J}=M_{0}d_{F}\frac{%
(2\pi T)\Delta ^{2}}{(\Delta ^{2}+\omega ^{2})^{3/2}}\; ,  \label{16}
\end{equation}%
where $M_{0}=\nu _{F}\mu _{B}J$ is the magnetic moment in F in the absence
of the proximity effect, {and we used the relation} $\sigma _{F}/\sigma
s=\nu _{F}D_{F}/\nu _{S}D_{S}$. At low temperatures $T\ll \Delta$, the
summation in Eq.(\ref{14}) is transformed into integration over $\omega $
that gives\cite{BVEepl04}
\begin{equation}
\mathcal{M}_{S}=-M_{0}d_{F}=-\mathcal{M}_{F}\; .  \label{17}
\end{equation}
This means that in the limiting case considered here the total magnetic
moment of Cooper pairs induced in the S region compensates the magnetic
moment of the F film\cite{footnoteM} If the condition (\ref{15}) is not
fulfilled or the temperature is not low enough the screening is not complete
and $|\mathcal{M}_{S}|<|\mathcal{M}_{F}|$. In the other limiting case of a
large exchange field $J\gg \epsilon _{bF}$ the induced magnetic moment in S
is given by
\begin{equation}
\mathcal{M}_{S}=-M_{0}d_{F}\epsilon _{bF}^{2}/2{J}^{2}\ll \mathcal{M}_{F}\; ,
\label{18a}
\end{equation}%
and therefore the screening is very weak.

The magnetic moment $M_{S}(0)$ induced right of the S-F interface can be
calculated from Eq.(\ref{16}) in the limit of low temperatures when the
function $m_{S}(\omega )$\ is approximated 
\begin{equation}
m_{S}(\omega )\cong M_{0}\frac{d_{F}}{\xi _{S}}\frac{2\pi T}{\Delta \lbrack
1+(\omega /\Delta )^{2}]^{5/4}} \; ,  \label{17a}
\end{equation}%
with $\xi _{S}^{2}=D_{S}/\sqrt{2}\Delta $.
 We then obtain
\begin{equation}
M_{S}(0)=-M_{0}\frac{d_{F}}{\xi _{S}}c_{0}  \label{17b}
\end{equation}%
with $c_{0}=\int_{0}^{\infty }dt(1+t^{2})^{-5/4}\cong 1.18$.

The magnetization in the F film can be written in the form $%
M_{F}=M_{0}+\delta M_{F}$, where $M_{0}$ is the uniform magnetization of the
ferromagnet in the absence of the proximity effect and $\delta M_{F}$ is a
correction due to the proximity effect.  As we consider a thin F layer with $%
d_{F}\ll \xi _{F}\approx \sqrt{D_{F}/J}$, the correction $\delta M_{F}$ can
be assumed constant in space. It can be shown that $\delta M_{F}$ is
negative, which leads to a decrease of the magnetization of the F film\cite%
{BVE04}. However, in what follows we neglect $\delta M_{F}$ since it does 
not affect qualitatively the main results.

We note that the condition, Eq. (\ref{15}), can be fulfilled in experiments%
\textbf{\ }with weak ferromagnets, as for example in Nb/CuNi structures as
those used in Ref.\cite{RyazanovPRL06}. By taking\textbf{\ }$R_{b}=R\cdot
L_{y}L_{z}\cong 30\mu \Omega \cdot 10\times 10\mu m^{2}$, $d_{F}\cong 20\dot{%
A}$, $D_{F}\cong 5cm^{2}/s$ and $\rho _{F}\cong 60\mu \Omega \cdot cm$ , we
obtain $d_{F}\rho _{F}/R_{b}\cong 0.4$ and $\epsilon _{bF}=D_{F}\rho
_{F}/R_{b}d_{F}=0.4D_{F}/d_{F}^{2}\cong 1200K$. For these parameters the
condition in Eq.\textbf{(}\ref{15}\textbf{) } is satisfied provided the
energy $J$ is not too large 


In this section we analyzed the proximity effect on the magnetic moment
induced in S. In the next section we find the spatial distribution of the
Meissner currents and magnetic fields in the whole S-F bilayer.

\section{\protect\bigskip Magnetostatics of a S-F bilayer}

In this section we determine the currents and fields induced in the S/F
structure shown in Fig. \ref{fig:geom}. The total current consists of two
contributions: the Meissner contribution and the current stemming from the
finite magnetization in the system. 
The magnetic induction $\mathbf{B}$ and the magnetic field $\mathbf{H}$ obey
the Maxwell equation both in the F and S films
\begin{eqnarray}
\mathbf{\nabla }\times \mathbf{B} &\mathbf{=}&\frac{4\pi }{c}\mathbf{j}+4%
\mathbf{\pi \nabla }\times \mathbf{M}  \label{2M} \\
\mathbf{\nabla }\times \mathbf{H} &\mathbf{=}&\frac{4\pi }{c}\mathbf{j}
\label{2aM}
\end{eqnarray}%
where $\mathbf{j}$ is the Meissner current denoted as $\mathbf{j}_{S,F}$ in
S or F films, respectively. In the F film, the current $\mathbf{j}_{F}$ is
carried by Cooper pairs induced due to the {proximity effect}. Both
currents $\mathbf{j}_{S,F}$ are related to the vector potential $\mathbf{A}%
_{S,F}$ via the London equation:
\begin{equation}
\mathbf{j}_{S,F}=-\frac{c}{4\pi }\frac{1}{\lambda _{S,F}^{2}}\mathbf{A}%
_{S,F}\;,  \label{3M}
\end{equation}%
where $\lambda _{S,F}$ is the London penetration length. We neglect
variation of $\lambda _{S,F}$ due to the proximity effect and assume that it
is constant. Moreover, in case of superconducting films with a short mean
free path $l$ the coherence length $\xi _{S}$ is usually much smaller than $%
\lambda _{S}$, which is equivalent to the limit of a  large Ginzburg-Landau
parameter $\kappa _{G-L}$, \
\begin{equation}
\kappa _{G-L}^{-2}=(\xi _{S}/\lambda _{S})^{2}=e^{2}nl^{2}/mc^{2}\ll 1,
\label{1a}
\end{equation}%
where $n$, $m$ and $l$ are the the concentration of free carriers, effective
mass and mean free path, respectively. For typical values of $n$ and $m$ one
obtains {as} upper limit for the mean free path is {\ $l\lesssim 1000%
\dot{A}$. }

As follows from Eqs.(\ref{2M}-\ref{3M}) the vector potential $\mathbf{A}%
_{S,F}$ satisfies the equation
\begin{equation}
\nabla ^{2}\mathbf{A}_{S,F}-\frac{1}{\lambda_{S,F}^{2}}\mathbf{A}%
_{S,F}=-4\pi \mathbf{\nabla }\times \mathbf{M}  \label{4M}
\end{equation}

In what follows we solve Eq.(\ref{4M}) for two different cases: in-plane and
out-of-plane orientation of $\mathbf{M}_{0}$.

\subsection{In-plane magnetization}\label{sec:inplane}

In this case $\mathbf{M}_{0}||\mathbf{e}_{z}$ and $\mathbf{B||M||e}_{z}$ and
$\mathbf{A||e}_{y}$. This menas that $\mathbf{B=(}0,0,B\mathbf{)}$ and $%
\mathbf{A=(}0,A,0\mathbf{)}$. {Then,} Eq. (\ref{4M}) reduces to
\begin{equation}
\partial _{xx}^{2}A_{S,F}-\frac{1}{\lambda _{S,F}^{2}}A_{S,F}=4\pi \partial
_{x}M_{S,F}  \label{5M}
\end{equation}%
As mentioned above, in the thin F film the magnetization is assumed to be
almost constant and therefore one can neglect the r.h.s of the previous
equation in the F region. The solution of Eq. (\ref{5M}) in  the F layer  {within the
limit }$d_{F}\ll \lambda _{F}$ can be written as
\begin{equation}
A_{F}(x)=a_{0}\left( 1+\frac{1}{2}\frac{x^{2}}{\lambda _{F}^{2}}\right)
+(h_{0}+4\pi M_{0})x\left( 1+\frac{1}{6}\frac{x^{2}}{\lambda _{F}^{2}}\right)
\label{8M}
\end{equation}%
where the coefficients $a_{0}$ and $h_{0}$ are integration constants.

Whereas in the S region the equation for $A$ is obtained by using Eq. (\ref%
{13a}) for the $M_S(x)$   induced in S:
\begin{equation}
\partial _{xx}^{2}A_{S}-\frac{A_{S}}{\lambda _{S}^{2}}=-4\pi \sum_{\omega
\geq 0}m_{S}(\omega )\kappa _{\omega }\exp (x\kappa _{\omega })\; .
\label{10M}
\end{equation}%
As demonstrated in the previous section the induced magnetization
(right-hand side of Eq. \ref{10M}) decays on a length of the order $\xi _{S}$%
. Since we consider the case $\xi _{S}\ll \lambda _{S}$, the solution in the
superconductor can be written as
\begin{equation}
A_{S}(x)=a_{S}\frac{\cosh [({x+d_{S}})/{\lambda _{S}}]}{\cosh \theta _{S}}%
-4\pi \sum_{\omega \geq 0}\frac{m_{S}(\omega )}{\kappa _{\omega }}[1+\delta
_{S}^{2}]\exp (x\kappa _{\omega })\text{,}  \label{11bM}
\end{equation}%
where $a_{S}$ is a third integration constant, $\delta _{S}=1/\lambda
_{S}\kappa _{\omega }$ is a small parameter ({see Eq.}(\ref{1a})), and $%
\theta _{S}=d_{S}/\lambda _{S}$. From Eq. (\ref{11bM}), one can obtain the
expressions for $B_{S}(x)=\partial _{x}A_{S}(x)$ and $H_{S}(x)=B_{S}(x)-4\pi
M_{S}(x)$, as shown in Appendix \ref{appendix:b}.

The integration constants $a_{0}$, $b_{0}$, and $a_{S}$ in Eqs.(\ref{8M}, %
\ref{11bM}) are determined by the following boundary conditions
\begin{equation}
\lbrack A]|_{x=0}=0\text{, }[H]|_{x=0}=0\text{, }H_{S}(-d_{S})=0\text{, }%
H_{F}(d_{F})=H_{ex}\text{. }  \label{12M}
\end{equation}%
where $[A]|_{x=0}\equiv A_{F}(0+)-A_{S}(0-)$. The first and second equations
provide the continuity of the vector potential $A$ and the field $H$ at the
interface. The condition assumes the presence of an external magnetic field
but in what follows we assume that $H_{ext}=0$.

In the main approximation we find three coupled equations determining $a_{s}$%
, $a_{0}$ and $h_{0}$. Their solution is given by
\begin{eqnarray}
a_{S} &=&-4\pi \frac{M_{0}\lambda _{S}\theta _{F}^{2}/2+\mathcal{M}%
_{S}\theta _{F}\lambda _{S}/\lambda _{F}-\Lambda }{\mathcal{D}_{S}}
\label{15M}\\
a_{0} &=&a_{S}+4\pi \mathcal{M}_{S}\text{,}  \label{13M} \\
h_{0} &=&\frac{a_{S}}{\lambda _{S}}\tanh \theta _{S}-4\pi \frac{\Lambda }{%
\lambda _{S}}\   \label{13bM} \; ,
\end{eqnarray}%
where $\theta _{F,S}=(d/\lambda )_{F,S}\ll 1$, $D_{S}=\tanh \theta
_{S}+\theta _{F}\lambda _{S}/\lambda _{F}$, and $\Lambda =\sum m_{S}\delta
_{S}^{2}\lambda _{S}$.  In deriving these equations we have used expression Eq. (\ref{14}) for the total
magnetisation $\mathcal{M}_{S}$ induced in the S region. %
%
%

Before analyzing the full spatial solution of the boundary problem let us
focus on the value of vector potential at the outer interface, $x=-d_{S}$.
The expression can be straightforwardly obtained from the above equation and reads:
\begin{equation}
A_{S}(-d_{S})=-4\pi \frac{\mathcal{M}_{S}\theta _{F}\lambda _{S}/\lambda
_{F}+(\theta _{F}^{2}/2)M_{0}\lambda _{S}-\Lambda }{\cosh \theta _{S}%
\mathcal{D}_{S}}\;,  \label{ASds}
\end{equation}%
From this expression one can already draw important conclusions regarding
the vector potential and    supercurrents, at large distances
from the boundary.  Let us consider two cases:

Case a): If one neglects the inverse proximity effect as done in Ref. \cite%
{Buzdin18}, the first and third term in the numerator of Eq. (\ref{ASds})
are zero and one obtains ($\mathcal{M}_{S}=0$)
\begin{equation}
A_{S,a}(-d_{S})\approx -4\pi M_{0}\lambda _{S}\frac{\theta _{F}^{2}}{2\sinh
\theta _{S}}\;.  \label{AM0}
\end{equation}%
{The corresponding magnetic induction at large distances from the S/F
interface coincides with Eq. (\ref{I5}) }

{Case b): If one takes into account the inverse proximity effect then
the second contribution anticipated in Eq. (\ref{I4a}) appears. }
Specifically in the full screening situation ($\mathcal{M}_{S}=-M_{0}d_{F}$%
), one obtains
\begin{equation}
A_{S,b}(-d_{S})=4\pi M_{0}d_{F}\frac{\gamma _{S}+\theta _{F}\lambda
_{S}/\lambda _{F}}{\sinh \theta _{S}}\;,  \label{AMfull}
\end{equation}%
where $\gamma _{S}=(c_{2}\kappa _{S}\xi _{S})=0.85(\xi _{S}/\lambda _{S})\ll
1$ and $c_{2}=\int_{0}^{\infty }dt(1+t^{2})^{-7/4}\approx 0.85$.
Clearly these two limiting cases describe very different situations, in
which the spontaneous currents have even different signs. It is important to
emphasize that even though the magnetization induced in the S layer occurs
over  the coherence length $\xi _{S}\ll\lambda_S$, it changes drastically {the} vector potential at
distances of the order of $\lambda _{S}$.

%
%
%
%
%
%
%

From the knowledge of the vector potential one can write the current density
$j(x)$ using the London equation, Eq.(\ref{3M}), as $j(x)=-(c/4\pi
)A(x)/\lambda _{S,F}^{2}$. The total currents through the F and S layers is
then defined as
\begin{equation}
I_{S,F}=\int_{S,F}dxj_{S,F}\;.  \label{N4}
\end{equation}%
From Eqs.(\ref{8M},\ref{11bM}) and in the leading order in our approach we
find that
\begin{equation}
\frac{4\pi }{c}I_{F}\approx -4\pi M_{0}\frac{\theta _{F}^{2}}{2}-4\pi
\mathcal{M}_{S}\theta _{F}^{2}/d_{F}  \label{N4a}
\end{equation}%
and
\begin{equation}
\frac{4\pi }{c}I_{S}\approx 4\pi M_{0}\frac{\theta _{F}^{2}}{2}+4\pi
\mathcal{M}_{S}\theta _{F}^{2}/d_{F}\;.  \label{N4b}
\end{equation}%
As there is no external field the currents $I_{S,F}$ sums to zero, $I_{F}+$ $%
I_{S}=0$. Remarkably, in the two limiting cases, a) and b), the total
current in the S and F films has a different sign, i.e. $I_{S,a}=-I_{S,b}=2%
\pi M_{0}\theta _{F}^{2}$.

%

Ff $J$ is large enough $J\gg \epsilon _{bF}$ there is a transition from
positive to negative $A_{S}(-d_{S})$ determined by a critical $J_c$. Indeed,
we obtain from Eqs.(\ref{18a})
\begin{equation}
A_{S}(-d_{S})=\frac{-2\pi M_{0}\theta _{F}^{2}\lambda _{S}}{\mathcal{D}%
_{S}\cosh \theta _{S} }\left[1-(\frac{J_{c} }{J})^{2}\right]  \label{N3c}
\end{equation}%
with $J_{c}=\epsilon _{bF}(\lambda_F/\lambda_S)\sqrt{c_2(\xi_S/d_{F}})$.

\begin{figure}[tbp]
\includegraphics[width=0.5\columnwidth]{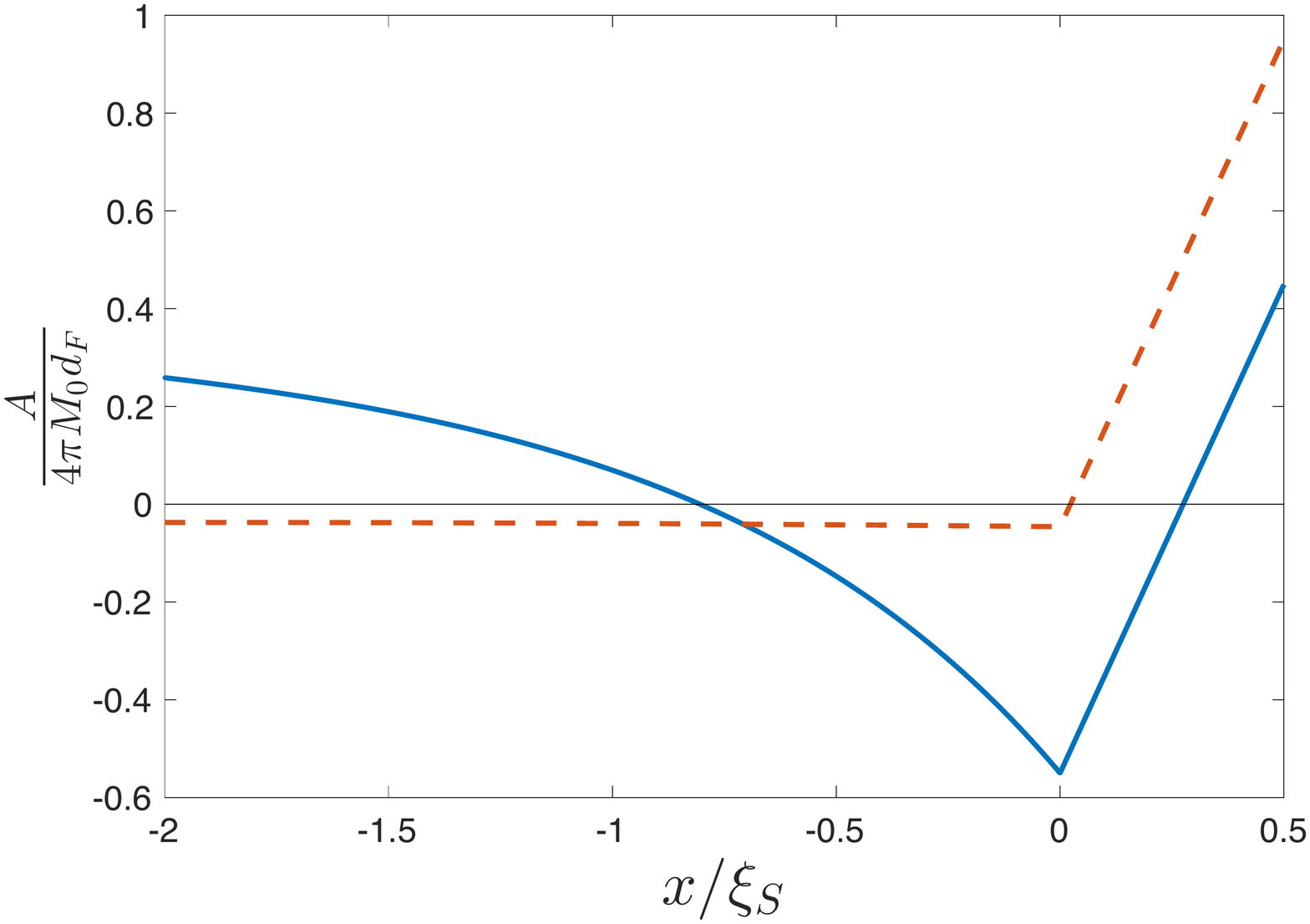} %
\includegraphics[width=0.5\columnwidth]{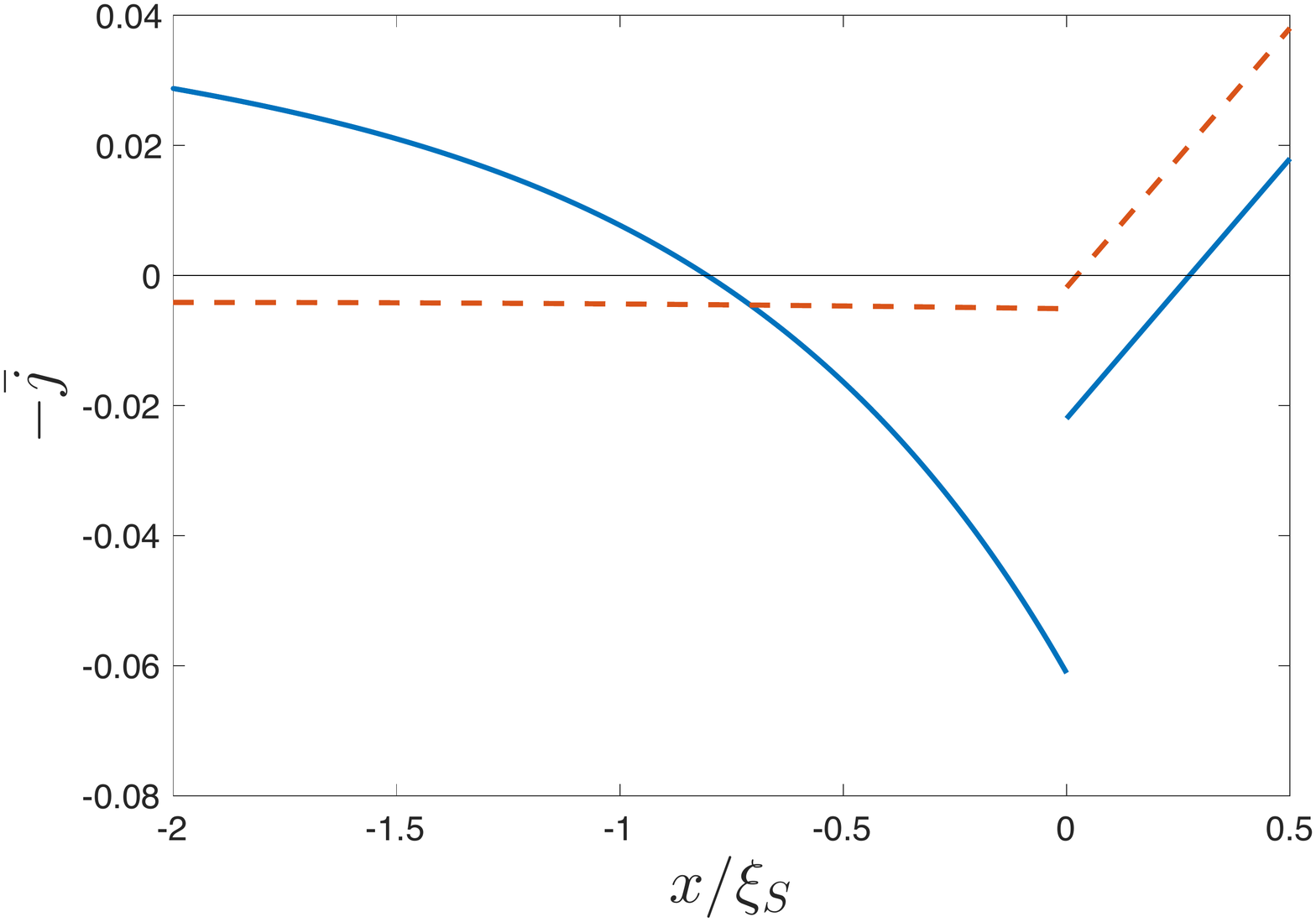}
\caption{(Color online.) Spatial dependence of the vector potential $A$
(upper panel) and the spontaneous current $\bar j\equiv j A\protect\xi%
_S^2/cM_0d_F$(lower panel). The solid line represents the case which
accounts for the inverse proximity effect (case b in the text), whereas the
dashed line shows the behaviour when the inverse proximity effect is
neglected (case a). Other parameters are chosen as $\protect\epsilon%
_b=5\Delta$, $d_S=2\protect\xi_S$, $d_F=0.5\protect\xi_S$, $\protect\kappa%
_{Sb}\protect\xi_S=0.1$, $\protect\kappa_{Fb}\protect\xi_S=1$ and $\protect%
\lambda_{F}=(5/3)\protect\lambda_{S}=5\protect\xi_S$. }
\label{fig:IAx}
\end{figure}


{In Fig. \ref{fig:IAx} we compare the spatial dependence of $A_{S}(x)$
(upper panel) and of the current density $j(x)$ induced in the system (lower
panel), in the two cases, a) no spin polarisation; $M_{S}=0$ , and b) when
the inverse proximity effect is taken into account. One clearly sees
qualitative differences between these two cases. In case b) the spontaneous
currents change sign at certain distance from the interface of the order of $%
\xi _S$, whereas in case a) the sign of the current in S is constant. In
addition, the amplitude of the spontaneous currents, and hence of the
magnetic inductance Eq. (\ref{I4a}), generated far from the S/F interface is
much larger in case b). These results demonstrate that for a correct
interpretation of experiments the induced magnetization in the
superconductor cannot be simply neglected. } 

%
%

\subsection{Out-of-plane magnetization}

Finally in this section we consider the case of a F layer with out-of-plane
magnetization: $\mathbf{M}_{0}||\mathbf{e}_{x}$. The field $\mathbf{H}%
_{S}(x)=(H_{S}(x),0,0)$ with $H_{S}(x)$ is determined from the equation $%
\mathrm{div} B_{S}(x)=0$:
\begin{equation}
\partial _{x}H_{S}(x)=-4\pi \partial _{x}M_{S}(x)\; .  \label{M15}
\end{equation}%
Thus we obtain
\begin{equation}
H_{S}(x)=-4\pi M_{S}(x)\, .  \label{M15a}
\end{equation}%
Equation $\mathbf{\nabla }\times \mathbf{H}_{S}(x)=(4\pi /c)\mathbf{j}$ then
yields
\begin{equation}
\mathbf{j}=0  \label{M16a}
\end{equation}%
The magnetic induction $\mathbf{B}$,  does not depend on $x$ and equals zero
both in the S and F films. 

\section{Conclusions}

We have studied the spatial dependence of the Meissner currents $j_{S}(x)$
and magnetic induction $B_{S}(x)$ that spontaneously arise in an S/F
structure even in the absence of an external field
. The fields $B_{S}(x)$ and $H_{S}(x)$ {originate} due to the orbital
and spin polarization effects and contain long-range and short-range
components (see Eq.(\ref{I4})). The amplitude of the short-range component $%
B_{s-r}(0)$ is due to the inverse {proximity effect} and is much
larger than $B_{l-r}(0)$.  {On the other hand}, the amplitude of the
long-range component $B_{l-r}(0,J)$ is caused by both, the  Meissner currents and
the spin polarization,  and depends on the magnitude of the exchange energy $J$ in the
F film.  It changes sign at $J\sim \epsilon _{bF}=D_{F}/(R_{b}\sigma
_{F}d_{F})$ being negative for $J\gg \epsilon _{bF}$ and positive for $%
J<\epsilon _{bF}$. Note that at large $J$\ the field $B_{S}$\ and $H_{S}$\
are small because both spin polarization (see Eq.(\ref{18a})) and orbital
effects are small. In Ref. \cite{Buzdin18} the inverse {proximit
effect} was neglected and therefore only the orbital contribution $%
B_{orb}(0)=-4\pi M_{0}\theta _{F}^{2}/2$ was obtained, where $\theta
_{F}^{2}=(d_{F}/\lambda _{F})^{2}\ll 1$. However, as explained above,  by
decreasing the exchange energy $J$, the inverse {proximity effect}  prevails, 
leads to a finite magnetic moment $%
\mathcal{M}_{S}\neq 0$,  and to a change of sign of  $B_{l-r}(0)$
changes sign. In such a case  its magnitude clearly  exceeds the value of $|B_{orb}(0)|$. 
Moreover,  also the results for the vector potential $A(x)$, and hence for the current
density  $j(x)$,  depend crucially on  the inverse proximity effect  (case (b) in section \ref{sec:inplane})
 and are qualitatively  different to the case in which 
this effect is neglected (case (a) in section \ref{sec:inplane} and Ref. \cite{Buzdin18}).
In particular, we find that the
Meissner current density $j_{S}$  changes sign in the S region if the induced magnetization is taken into account.

\section*{Acknowledgements}

F.S.B acknowledges financial support from Horizon research and innovation
programme under grant agreement No. 800923 (SUPERTED) the Spanish Ministerio
de Econom\'{\i}a, Industria y Competitividad (MINEICO) under Project
FIS2017-82804-P.

\appendix 

\section{Solution of the Usadel equation in the S-F strcuture}

\label{app_usadel}

The Usadel equations have the form

\begin{equation}
-D_{S}\partial _{x}(\hat{g}_{S}\partial _{x}\hat{g}_{S})+\omega \lbrack
\tau_3,\hat{g}_{S}]+\Delta \lbrack \tau_1,\hat{g}_{S}]=0\text{, S film}
\label{A3}
\end{equation}

\begin{equation}
-D_{F}\partial _{x}(\hat{g}_{F}\partial _{x}\hat{g}_{F})+\omega \lbrack
\tau_3,\hat{g}_{F}]+iJ[ \tau_3\sigma_3,\hat{g}_{F}]=0\text{, F film}  \label{A4}
\end{equation}%
Eqs.(\ref{A3}-\ref{A4}) are complemented by the normalization relation

\begin{equation}
(\hat{g}\cdot \hat{g})=\hat{1}  \label{A4a}
\end{equation}%
and the boundary conditions (\cite{KupLukichev88})
\begin{equation}
(\hat{g}\partial _{x}\hat{g})_{F}=\kappa _{bF}[\hat{g}_{S},\hat{g}_{F}]\text{%
, }(\hat{g}\partial _{x}\hat{g})_{S}=\kappa _{bS}[\hat{g}_{S},\hat{g}_{F}]
\label{A5}
\end{equation}%
where $\kappa _{bF(S)}=(R_{b}\sigma _{F(S)})$ $^{-1}$, $R_{b}$ is the
interface resistance per unit area.

The lineraised Eq.(\ref{A3}) is
\begin{equation}
-\partial _{xx}^{2}\delta \hat{g}_{S}+\kappa_\omega^{2}\delta \hat{g}_{S}=2\delta
\Delta \omega (\omega \tau_1-\Delta \tau_3)/D_{S}  \label{A10}
\end{equation}%
where $\kappa_\omega^{2}=2\sqrt{\omega ^{2}+\Delta ^{2}}/D_{S}$. We used the
relation

\begin{equation}
\delta \hat{g}_{S}\cdot \hat{g}_{S}+\hat{g}_{S}\cdot \delta \hat{g}_{S}=0
\label{A10a}
\end{equation}%
which follows from the normalization condition, Eq.(\ref{A4a}).

The induced magnetization is determined by the component $\delta g_{S3}=$Tr$%
(\tau_3\hat{g}_{S})/4$ (see Eq.(\ref{6a})). We multiply Eq.(\ref{A10}) by $%
\tau_3\sigma_3$ and calculate the trace. We find the solution
\begin{equation}
g_{S3}(x)=g_{S3}(0)\exp (x\kappa_\omega)  \label{A11}
\end{equation}
The integration constant is found from the boundary condition Eq.(\ref{A5})
that yields
\begin{equation}
\partial _{x}g_{S3}(x)|_{x=0}=2\kappa _{Sb}F_{S}[F_{S}g_{F3}-G_{S}f_{F3}]
\label{A12}
\end{equation}%
where $f_{F3}=\epsilon _{bF}F_{S}$Im$(1/\tilde{\zeta}_{\omega +})$ and $%
g_{F3}=$Im$(\tilde{\omega}_{+}/\tilde{\zeta}_{\omega +})$. The function $%
\tilde{\zeta}_{\omega +}$ is defined in Eq.(\ref{8}).

We obtain for $g_{S3}(x)$
\begin{equation}
g_{S3}(x,\omega )=\frac{2\kappa _{bS}}{\kappa}F_{S}[F_{S}g_{F3}-G_{S}f_{F3}]%
\exp (x\kappa_\omega)\equiv g_{33}^{(S)}(0,\omega )\exp (x\kappa_\omega)  \label{A13}
\end{equation}%
One can see that this correction $\delta \hat{g}_{S}$ is small if the
condition

\begin{equation}
\kappa _{bS}\xi _{S}\ll 1\text{ }  \label{A13c}
\end{equation}%
is fulfilled, that is, $R_{Sb}\gg \rho _{S}\xi _{S}$.

\section{The magnetic field}

\label{appendix:b}

For completeness we show in this appendix the expressions for the magnetic
induction $B(x)=\partial _{x}A(x)$ and magnetic field $H(x)=B(x)-4\pi M_{F}$
which can be obtained from Eqs.(\ref{8M},\ref{11bM})
\begin{eqnarray}
B_{F}(x) &\cong &a_{0}\frac{x}{\lambda_F^2}+(h_{0}+4\pi M_{0})(1+\frac{ x^{2}%
}{2\lambda_F^{2}})  \label{A8a} \\
H_{F}(x) &\cong &a_{0}\frac{x}{\lambda_F^2}+h_0\left(1+\frac{x^2}{%
2\lambda_F^2}\right)+4\pi M_0\frac{x^2}{2\lambda_F^2}  \label{A8b}
\end{eqnarray}

\begin{eqnarray}
B_{S}(x) &=&\frac{a_{S}}{\lambda_S}\frac{\sinh ((x+d_{S})/\lambda_{S})}{\cosh
\theta _{S}}+4\pi M_{S}(x)-4\pi \sum_{\omega \geq 0} m_{S}(\omega
)\delta_S\exp (x\kappa_\omega)  \label{A1} \\
H_{S}(x) &=&\frac{a_{S}}{\lambda_S}\frac{\sinh ((x+d_{S})/\lambda_{S})}{\cosh \theta
_{S}}-4\pi \sum_{\omega \geq 0}m_{S}(\omega )\delta_S^2\exp (x\kappa_\omega) \; ,
\label{A1a}
\end{eqnarray}
where $\delta_S=1/\lambda_S\kappa\omega$.

\bigskip


\begin{thebibliography}{99}
\bibitem{footnoteM} {Here and in the rest of the article, when
talking about spin screening, we refer to the screening of the magnetic
moment of free electrons (itinerant ferromagnet) which equals to $%
\mu_B\nu_FJ $. The contribution to the magnetization of F stemming from
localized magnetic moments is not screened by the inverse proximity effect%
\cite{BVE04}.}
\bibitem{foot} Note that induced magnetisation in S was also considered in a
ferromagnetic-insulator/superconductor structure \cite{Tokuyashu02,Koshina02}%
. Whereas in Ref. \cite{Koshina02} a diffusive S/F bilayer was analysed.
Contrary to our results, in the latter work the induced magnetization
obtained has the same orientation as the magnetic moment in the ferromagnet.
\end{thebibliography}
\end{document}